\documentclass[english,aps,prl,twocolumn,superscriptaddress,reprint]{revtex4-1}
\usepackage[T1]{fontenc}
\usepackage[utf8]{inputenc}
\usepackage{color}
\usepackage{babel}
\usepackage{fancybox}
\usepackage{calc}
\usepackage{textcomp}
\usepackage{amsmath}
\usepackage{amssymb}
\usepackage{graphicx}
\usepackage{wasysym}
\usepackage[unicode=true,
 bookmarks=true,bookmarksnumbered=false,bookmarksopen=false,
 breaklinks=false,pdfborder={0 0 0},pdfborderstyle={},backref=false,colorlinks=true]
 {hyperref}
\hypersetup{pdftitle={Band structure of a IV-VI black phosphorus analogue, the thermoelectric SnSe},
 pdfauthor={Ivo Pletikosić},
 pdfkeywords={SnSe, thermoelectric, black phosphorus analogue, ARPES},
 urlcolor=blue,linkcolor=black}
\begin{document}

\title{Band structure of a IV-VI black phosphorus analogue, the thermoelectric
SnSe}

\author{I. Pletikosić}
\email{ivop@princeton.edu}

\affiliation{Department of Physics, Princeton University, New Jersey, USA}

\affiliation{Condensed Matter and Materials Science Department, Brookhaven National
Laboratory, New York, USA}

\author{F. von Rohr}

\affiliation{Department of Chemistry, Princeton University, New Jersey, USA}

\author{P. Pervan}

\affiliation{Institut za fiziku, Zagreb, Croatia}

\author{P. K. Das}

\affiliation{Istituto Officina dei Materiali (IOM-CNR), Laboratorio TASC, Trieste,
Italy}

\affiliation{International Centre for Theoretical Physics, Trieste, Italy}

\author{I. Vobornik}

\affiliation{Istituto Officina dei Materiali (IOM-CNR), Laboratorio TASC, Trieste,
Italy}

\author{R. J. Cava}

\affiliation{Department of Chemistry, Princeton University, New Jersey, USA}

\author{T. Valla}

\affiliation{Condensed Matter and Materials Science Department, Brookhaven National
Laboratory, New York, USA}
\begin{abstract}
The success of black phosphorus in fast electronic and photonic devices
is hindered by its rapid degradation in presence of oxygen. Orthorhombic
tin selenide is a representative of group IV-VI binary compounds that
are robust, isoelectronic, and share the same structure with black
phosphorus. We measured the band structure of SnSe and found highly
anisotropic valence bands that form several valleys having fast dispersion
within the layers and negligible dispersion across. This is exactly
the band structure desired for efficient thermoelectric generation
where SnSe has shown a great promise.  
\end{abstract}

\date{\today}

\maketitle
A growing interest in tin selenide emerged with the reporting of an
unusually high thermoelectric figure of merit $zT$ of 2.6 \citep{ZhaoNAT,ZhaoSCI}.
SnSe has also been considered as a prospective material for ultrathin
photovoltaic films \citep{Franzman} and Li-ion battery anodes \citep{Lee}.
 In addition, SnSe happens to be a group IV–VI (Sn: 5s\textsuperscript{2}5p\textsuperscript{2},
Se: 4s\textsuperscript{2}4p\textsuperscript{4}) isoelectronic analogue
of black phosphorus of group V (P: 3s\textsuperscript{2}3p\textsuperscript{3}).
The sheets building black phosphorus, called phosphorene, have already
shown a great promise in fast-response electronic and photonic devices
\citep{BP_Koenig,BP_Xia,BP_Li,BP_Liu,BP_Buscema,BP_Carvalho}. Sharing
the almost defect-free two-dimensional structure with its flat cousin
graphene, phosphorene displays relatively high charge carrier mobility
\citep{BP_Li,BP_Qiao,BP_Long}, but unlike graphene has an inherent
bandgap, which, combined with various contact materials to offset
the chemical potential, is essential for creating the on/off flow
of electrons in digital logic, in the generation of photons in LEDs
and lasers or photon absorption in photovoltaics. The band gap of
ultrathin black phosphorus films has also been shown to depend on
the thickness, and with the values of 0.3 to 2 eV, advantageously
bridges the zero bandgap of graphene and 1–2.5 eV gaps of transition
metal dichalcogenides. Unfortunately, the material degrades within
a few hours when exposed to oxygen and water vapor in air \citep{Huang}.
This is why related but stable layered semiconductors are of significant
interest. The subject of this study, SnSe, shares a similar creased
honeycomb layered crystal structure with black phosphorus, exhibits
even higher carrier mobility, up to several thousand cm\textsuperscript{2}V\textsuperscript{-1}s\textsuperscript{-1}
\citep{Nassary,Maier,DFT_Zhang}, and like many similar layered materials
can be  exfoliated or epitaxially grown to yield films of definite
atomic thickness \citep{Ju,LLi,Zhang,XLi}. SnSe, moreover, does not
deteriorate in air. All this makes it a desirable isostructural, isoelectronic
counterpart to black phosphorus. 

The calculated band structure of SnSe has been reported in numerous
density functional studies \citep{ZhaoSCI,DFT_Ding,DFT_Gomes,DFT_Gonzalez,DFT_Hong,DFT_Kutorasinski,DFT_Mori,DFT_Shafique,DFT_Shi,DFT_Sirikumara,DFT_Suzuki,DFT_Yu,DFT_Yang,DFT_RGuo},
with no experimental verification so far. Here, we report on features
in the band structure of low-temperature SnSe we find important for
the thermoelectric transport: pronounced anisotropy that results in
different response along the three crystal axes, valence band consisting
of multi-valley hole pockets, that are in turn capable of supporting
highly mobile, low-effective-mass charge carriers. As some of these
have not been precisely captured in existing \emph{ab initio} simulations,
our results will lead to a better understanding of both low and high-temperature
phases of this intriguing material. 

ARPES measurements were conducted at the APE beamline of Elettra Sincrotrone
Trieste using a Scienta DA30 electron analyzer in the lens deflection
mode that enables motion-free recording of the polar and tilt emission
angles in the cone of a 30° opening. The azimuthal angle of the sample
was fixed at the time of the mounting to the holder. The resulting
momentum-space maps have been rotated by 22° to conveniently align
the high-symmetry axes; the parts left out of the 30° cone were reconstructed
by symmetrization. Photons in the range of 30–50~eV were used for
the photoexcitation. The chemical potential of semiconducting samples
was set to the position of the Fermi level ($0=E_{F}$) of a nearby
metallic sample. The temperature was held at 25~K. 

SnSe was synthesized in an evacuated quartz tube from 5N tin shot
and polished 3N5 selenium shot at 950 °C for 48h. Single crystals
were obtained by the vertical Bridgman method from a 950 °C melt of
the fine powder of the resulting material enclosed in a 5 mm quartz
tube that was drawn through the furnace at 0.2 mm/h. 

\begin{figure}
\includegraphics{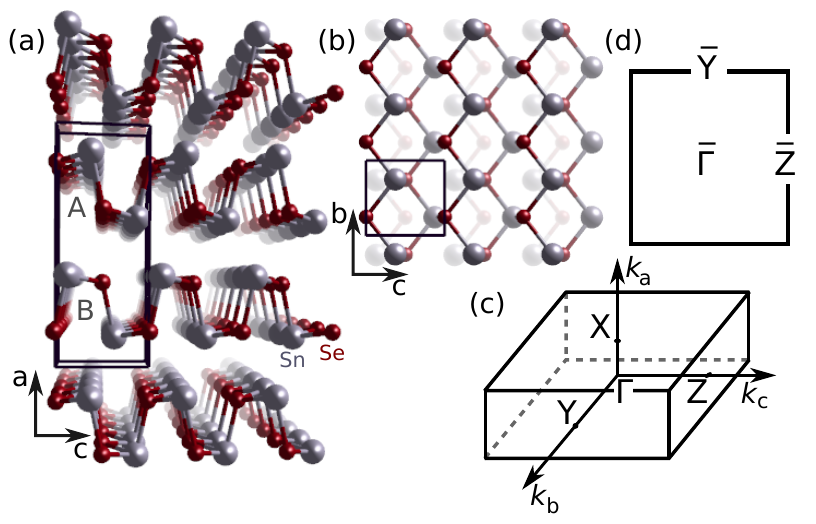}

\caption{\label{fig:Crystal_structure}The crystal structure of SnSe viewed
(a) along, and (b) perpendicular to the layers. The lower of the two
layers contained in the unit cell, A and B, has in panel (b) been
made nearly invisible. The unit cell size is 11.5\texttimes 4.2\texttimes 4.4~Å\protect\textsuperscript{3}.
The reciprocal space repetition cells: (c) the three-dimensional bulk
Brillouin zone with $\Gamma$X=0.27~Å\protect\textsuperscript{-1},
$\Gamma$Y=0.75~Å\protect\textsuperscript{-1}, $\Gamma$Z=0.71~Å\protect\textsuperscript{-1};
(d) the layer-bound two-dimensional surface Brillouin zone. The points
Y and Z project along the $a$ direction into $\bar{Y}$ and $\bar{Z}$,
respectively.}

\end{figure}

The crystal of SnSe consists of loosely bound  layers that are, in
the low temperature phase (T<800~K), built as an intricate knitting
of tin and selenium atoms, in which each atom is linked to three of
the other species by covalent bonds in trigonal pyramid geometry,
forming distinct ridges along one of the crystal directions; see Figs.
1(a) and 1(b). The structure is such that a single layer, stretched
into a flat sheet, would form a honeycomb lattice of alternating Sn
and Se atoms. Neighboring layers are related by a center of inversion,
but are lacking one themselves \citep{DFT_Shi}. 

\begin{figure}
\includegraphics{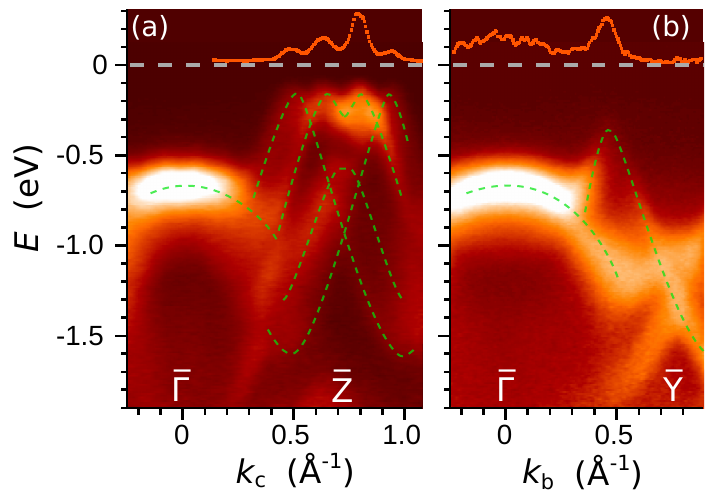}

\caption{\label{fig:Band_dispersion}Photoemission spectral function showing
the dispersion of the occupied electronic bands of SnSe along the
high-symmetry directions (a) $\bar{\Gamma}-\bar{Z}$, and (b) $\bar{\Gamma}-\bar{Y}$
of the surface Brillouin zone. Dashed curves are tracing the bands
that will be discussed in the text. The intensity profiles on top
were taken at (a) -120~meV and (b) -350 meV. Photon energy was $h\nu$=34~eV.}
\end{figure}

We start by describing the electronic band structure along two high-symmetry
directions of the in-plane momentum space. The photoemission spectra
recorded using $h\nu$=34~eV photons, displayed in Fig. \ref{fig:Band_dispersion},
exhibit a multitude of downward dispersing bands at $\bar{\Gamma}$,
from which several start rising  in the direction of $\bar{Z}$,
and one in the direction of $\bar{Y}$, reaching, however, somewhat
lower final energy before turning down again. The overlaid curves
explicate the full band configuration. This has been deduced from
a set of photon energy-dependent measurements, having the third component
of the crystal momentum $k_{a}$ span approximately one and a half
bulk Brillouin zones (see Supplemental Material \citep{Suppl}).

When it comes to thermal and electronic transport properties of a
material, they are in the valence band defined by the bands highest
in energy, thus closest to the chemical potential. These will be discussed
in more detail in what follows. 

\begin{figure*}
\includegraphics{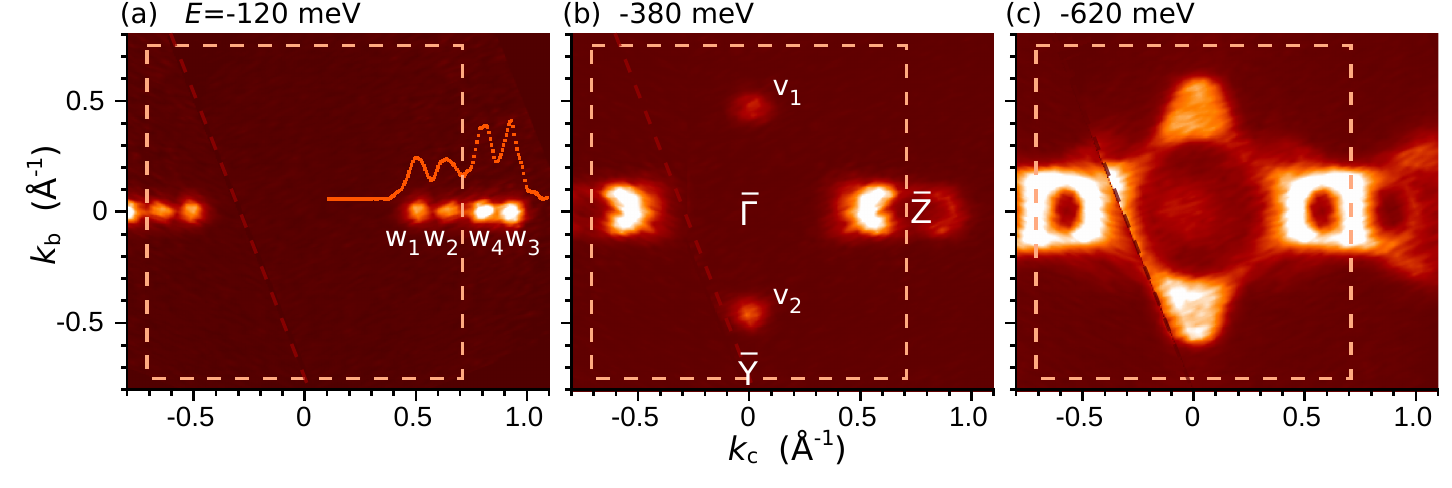}

\caption{\label{fig:Constant_energy_maps}ARPES intensity maps in the momentum
plane parallel to the layers of SnSe taken at the energies of (a)
-120~meV, (b) -380~meV, and (c) -620~meV. The lower left corner
of the maps was reconstructed by symmetrization from the opposite
side. Photons of $h\nu$=38~eV were used in (a); $h\nu$=50~eV in
(b) and (c). The dashed rectangles represent the surface Brillouin
zone. Pockets along $\bar{\Gamma}$–$\bar{Z}$ received labels $w_{1}$–$w_{4}$;
those along $\bar{\Gamma}$–$\bar{Y}$, $v_{1}$ and $v_{2}$. }
\end{figure*}

Figure \ref{fig:Constant_energy_maps} shows the momentum space configuration
of the highest lying bands in the occupied band structure. At -120~meV
four tiny collinear pockets, labeled $w_{1}\dots w_{4}$ start showing
along $\bar{\Gamma}$–$\bar{Z}$ in the in-plane cut of the bulk Brillouin
zone (Fig \ref{fig:Constant_energy_maps}a). The map was acquired
using 38~eV photons, representing the perpendicular momentum ($k_{a}$)
shift of about 0.15~Å\textsuperscript{-1} with respect to the spectra
in Fig. \ref{fig:Band_dispersion}. No shift of the band maxima can
be resolved either in momentum (they are found on either side, 0.08
and 0.22 Å\textsuperscript{-1} away from $\bar{Z}$) or energy (ARPES
intensity starts appearing at -120~meV). Only a change in relative
intensities is observed when the intensity profile in Fig. \ref{fig:Constant_energy_maps}(a)
is compared to the profile on top of Fig. \ref{fig:Band_dispersion}(a),
and is most likely due to a variation in optical transition probabilities.
Were the maximum energies that the neighboring bands $w_{1;3}$, $w_{2;4}$
reach different, their intensity profiles at constant energy would
inevitably differ as well; that is not the case here.  As the energy
is lowered, the circular sections of these bands grow in size and
soon merge. A gap of 250~meV opens up at $\bar{Z}$ marking the avoided
crossing of the bands $w_{2}$ and $w_{4}$. Interestingly, no such
gaps are observed at the crossings of the other two pairs—$w_{1}$,$w_{2}$
and $w_{4}$,$w_{3}$ in Figs. \ref{fig:Band_dispersion}(a), \ref{fig:Constant_energy_maps}(b)
or any of the figures in \citep{Suppl}.

Two other hole pockets, labeled $v_{1}$, $v_{2}$ in Fig. \ref{fig:Constant_energy_maps}(b),
emerge in the perpendicular direction, centered at $k_{b}$ =\textpm 0.45~Å\textsuperscript{-1}
with apices below some -380~meV. Lastly, the first photoemission
signatures of the circular bands around $\Gamma$ become visible just
below the energy of -600 meV. 

Together with almost negligible dispersion of the bands across the
material's layers, that we observe in several other $k_{a}$ cuts
using a range of photon energies and show in \citep{Suppl}, Fig.
\ref{fig:Constant_energy_maps} adds to the notion of the unusual
anisotropy of the band structure in all three crystal directions that
can have significant consequences in many transport properties. 

The effective mass and band velocity are the factors that decide the
charge carrier mobility. In Fig. \ref{fig:Pockets_fitting} we reexamine
the two highest lying hole pockets and determine the effective mass
by fitting the band dispersion to parabolas locally centered at pocket
maxima.

\begin{figure*}
\includegraphics{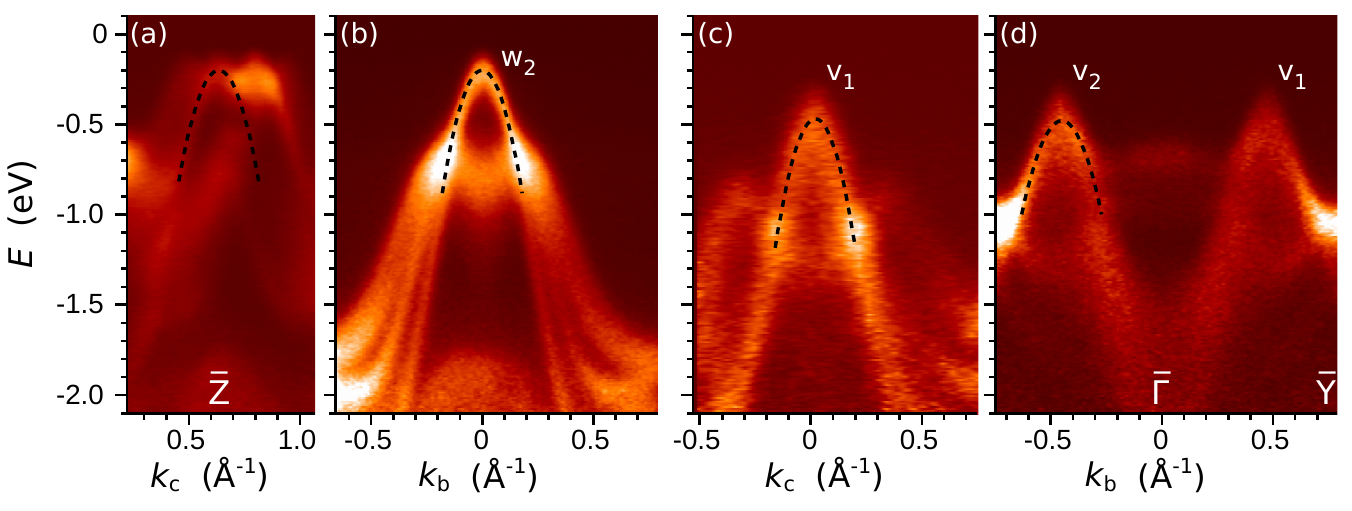}

\caption{\label{fig:Pockets_fitting}Energy dispersion cuts of the highest-lying
pockets in the valence band structure of SnSe along two perpendicular
directions in the momentum plane parallel to the layers. Panels (a-b)
show the holelike pockets centered around $\bar{Z}$; (c-d) the pockets
in the vicinity of $\bar{Y}$. The bands are overlaid by parabolic
fits used to estimate the effective mass. 34~eV photons were used
in (a-b); 50~eV in (c-d).}
\end{figure*}

Four hole pockets in the vicinity of $\bar{Z}$ appear to have very
similar initial dispersion along $\bar{\Gamma}$–$\bar{Z}$ direction,
Fig. \ref{fig:Pockets_fitting}(a). Fitting to a parabola $\frac{\hbar^{2}}{2m}(k_{c}-k_{w})^{2}$
for any of the apices $w$ gives for the 'effective mass' $m_{c}^{w}=0.21\,m_{e}$.
 The parabolic dispersion in the perpendicular direction (Fig. \ref{fig:Pockets_fitting}(b)
is a cut for the second of the four) is slightly steeper, giving for
the mass parameter of these pockets in the direction of $k_{b}$ the
value of $m_{b}^{w}=0.18\,m_{e}$. The fitting curves place the top
of the valence band at 200~meV below the chemical potential. The
pockets along the $\bar{\Gamma}$–$\bar{Y}$ direction are best visible
in ARPES when the excitation energy of 50~eV is used (this represents
a shift of 0.57\textpm 0.03 Å\textsuperscript{-1} in $k_{a}$, to
be compared to $2\frac{\pi}{a}$ of 0.54~Å\textsuperscript{-1}).
Their cuts in two perpendicular in-plane direction are shown in Figs.
\ref{fig:Pockets_fitting}(c) and \ref{fig:Pockets_fitting}(d). Parabolic
fitting places the centroid of the intensity at 480~meV below the
chemical potential. The pockets $v$ have the effective mass of $m_{c}^{v}=0.17\,m_{e}$
in the $c$ direction, and $m_{b}^{v}=0.24\,m_{e}$ perpendicular
to it. We show these summits and the fitting curves in greater detail
in Fig. S3 of \citep{Suppl}.

\vspace{0.2cm}

The key to the efficient generation of the electric power and the
record-high $zT$ of 2.6 in high temperature phase of SnSe (T>800~K)
is believed to be the unusually strong lattice anharmonicity and,
consequently, a very low lattice thermal conductivity $\kappa_{l}$
\citep{ZhaoNAT,Carrete,Li}. A similarly high $zT$ of 2.0 was reported
for p-doped single crystals in the low-temperature phase at 770~K.
We will demonstrate here that the band structure of low-temperature
SnSe has enough ingredients for a high thermopower, $PF=\sigma S^{2}$,
and together with inherently low thermal conductivity of the crystal
should support the efficient power generation or cooling, as assessed
by maximizing the figure of merit $zT=\frac{\sigma S^{2}}{\kappa_{\mathrm{e}}+\kappa_{l}}T$.
Unfortunately, the parameters in this equation available for such
optimization—the electronic conductivity $\sigma$, the Seebeck coefficient
$S$, and the electronic thermal conductivity $\kappa_{e}$, are interdependent
and mutually opposing. 

In order to achieve a high Seebeck voltage, the band structure around
the chemical potential needs to act as a filter that favors the heat
diffusion by either cold carriers below the chemical potential or
hot carriers above, and not both equally at the same time \citep{Datta}.
That means having the chemical potential in the region of rapidly
changing density of states, or better yet, in a semiconductor's gap
closer to one and farther from the other extremum of the conduction
and valence bands. With the predicted band gap of around 800~meV,
and the valence band maximum at $\apprle$200 meV below the chemical
potential, our SnSe samples appear naturally \emph{p}-doped, and thus
well suited for voltage generation. We note, however, that part of
the asymmetry at low temperatures is likely coming from unequal onsets
of density of states at two ends of the gap and may change as the
chemical potential shifts at higher $T$ \citep{DFT_Hong}. The higher
voltage, in general, will be generated by a temperature gradient the
farther in $kT$ (at 600K, $kT$=50~meV) the chemical potential is
from the extremum of the \emph{conducting} band (that is the valence
band in p-doped, or the conduction band in n-doped crystals). This
unfortunately leads to a small carrier density $n$, reduced conductivity
$\sigma=n\mu$, and severely lowers the available thermopower.  

 Tin selenide, as we have seen above, has in the low-energy occupied
band structure a few fast-dispersing bands capable of hosting highly
mobile carriers. Coupled with structural faultlessness, their high
mobility ($\mu$) can potentially balance out this reduced carrier
density, providing enough conductivity for sizable thermopower. 

Yet another feature works in favor of a high Seebeck coefficient in
SnSe: the total band dispersion across the layers is extremely small.
This implies a high effective mass $m_{a}$. Assuming a tight-binding
form of dispersion along $a$, $E(k_{b},k_{c})-\Delta(1-\cos ak_{a})$,
where $2\Delta$ is below our detection limit of some 30-50~meV,
we estimate $m_{a}$ to be $2.5\,m_{e}$ or higher. When a thermal
gradient is established along the layers, the uttermost value that
the figure of merit can attain is expressed by the formula which summarizes
the desired properties of an ideal thermoelectric material (Ref. \citep{Sootsman}):
\[
z_{\max}\propto\gamma\,\tau_{b|c}T^{3/2}\kappa_{l}^{-1}\sqrt{\frac{m_{c|b}}{m_{b|c}}m_{a}}
\]

The value is, fortuitously, proportional to $m_{a}$. When conduction
along either $b$ or $c$ high-symmetry directions is assumed (the
$|$ above stands for \emph{or}), the effective masses $m_{b}$, $m_{c}$
we determined to be of comparable size for a number of valence bands,
will in the above formula nearly cancel. Given the high anisotropy
of the band structure, it should not be excluded that the ratio of
the masses will increase when the two perpendicular in-plane directions
are allowed to stray. The extreme in-plane band anisotropy of Fig.
\ref{fig:Constant_energy_maps}, not found in most but a few two-dimensional
materials \citep{Pletikosic}, may be detrimental to the electronic
heat transport as it limits the number of configurations available
for the electrons to hop between in their random thermal motion. The
anisotropy is also likely to affect the value of the scattering time
$\tau$ of the carriers moving along the transport direction, and
should be considered when optimizing the figure of merit as well.
Interestingly enough, a recent angle-dependent conductivity study
of SnSe finds the hole conductivity along $b$ almost four times higher
than along $c$ \citep{Xu}. The quasiparticle effective masses on
top of the valence band, which we find almost equal in both in-plane
directions, cannot account for the observed anisotropy. Our toy-model
calculation, by the recipe of Ref. \citep{DFT_Gonzalez}, for a single
hole band having the dispersion of SnSe bands, indicates that the
scattering times along the two directions should differ by an order
of magnitude in order to reproduce the conductivity angular dependence
\citep{Suppl}. This, curiously, does not lead to a notable Seebeck
coefficient anisotropy and the power factor is only affected in the
first order, through $\sigma$. Our model also nicely shows the benefits
of having little out-of-plane dispersion: power factor maxima are
achieved for completely filled bands in the $k_{a}$ direction, and
increase as the bands narrow. 

Finally, the ingredient in the band structure of SnSe that can easily
lead to doubling the figure of merit is the band degeneracy factor
$\gamma$, which reflects the number of available modes in thermoelectric
transport. We have found four hole pockets along $\bar{\Gamma}$–$\bar{Z}$
that appear at exactly same energy in our constant energy maps, Fig.
\ref{fig:Constant_energy_maps}(a), and have the same initial parabolic
dispersion. This is an important finding that \emph{ab initio} thermoelectric
transport simulations should rely on.

Density functional studies of bulk SnSe \citep{DFT_Ding,DFT_Gomes,DFT_Gonzalez,DFT_Hong,DFT_Kutorasinski,DFT_Mori,DFT_Shafique,DFT_Shi,DFT_Sirikumara,DFT_Suzuki,DFT_Yu,DFT_Yang,DFT_RGuo}
generally agree on the indirect nature of the band gap that forms
between the valence band pockets along $\Gamma$–$Z$ and several
electron pockets in the conduction band showing along $\Gamma$–$Y$.
The electron pocket minima themselves, however, happen to be only
a few tens of meV apart. Most studies underestimate by some 200–300
meV the size of the gap that was experimentally determined to be 860~meV
\citep{ZhaoNAT}. They also underestimate the difference in energy
of the valence band pockets $w$ and $v$, that we establish at 280~meV,
the difference ranging from 0~meV in \citep{DFT_Hong,DFT_Sirikumara}
to 100–180~meV in \citep{ZhaoSCI,DFT_Suzuki,DFT_Mori,DFT_Kutorasinski,DFT_Shi,DFT_Gomes,DFT_Gonzalez,DFT_Yang,DFT_RGuo}.
The discrepancy is likely to affect the existing predictions of transport
properties at any higher temperatures. There is also a disagreement
in the number and nature of the pockets along $\Gamma$–$Z$: while
majority of studies show four, the bands not always reach the same
energy, and the 250~meV gap at $\bar{Z}$ due to the avoided crossing
of the bands $w_{2}$ and $w_{4}$ never exists; most studies have
it, however, between the pairs $w_{1}$,$w_{2}$ and $w_{4}$,$w_{3}$,
where we, interestingly, observe none. Similar discrepancies exist
for the effective masses, where calculations generally show much higher
variability than we have observed. \emph{Ab initio} simulations tend
to ascribe the anisotropy of the transport properties to the difference
in the effective masses, mostly disregarding the anisotropy of the
scattering time \citep{Xu}. This all emphasizes the complexity of
the band structure near band extrema and sensitivity of band structure
calculations that we are hoping to steer in the right direction. 
Our results will also help modeling and exploration of ultrathin layers
of SnSe in search for advantageous properties already found in black
phosphorus—the band gap variable in size and (in)directionality \citep{DFT_Guo,DFT_Kamal,DFT_Wang,DFT_Zhang},
or enhancements of thermoelectric properties by band engineering and
doping \citep{Pei,DFT_Wang,ZhaoSCI,Duong,DFT_Shafique,Gharsallah,DFT_Zhu},
spin dependent transport in a single layer \citep{DFT_Shi}, or valleytronics
by selective optical pumping from the valence to the conduction band
valleys using linearly polarized light \citep{Rodin}. 
\begin{acknowledgments}
 This work was supported by the US Department of Energy, Contract
No. DE-SC0012704, the ARO MURI program, Grant No. W911NF-12-1-0461,
and was partly performed at the Nanoscience Foundry and Fine Analysis
(NFFA-MIUR Italy, Progetti Internazionali) facility. The crystal growth
at Princeton University was supported by the Gordon and Betty Moore
Foundation, EPiQS initiative, grant GBMF-4412. 
\end{acknowledgments}

\cleardoublepage{}

\onecolumngrid
\raggedbottom

\appendix

\title{Band structure of a IV-VI black phosphorus analogue, the thermoelectric
SnSe}

\author{I. Pletikosić}

\affiliation{Department of Physics, Princeton University, New Jersey, USA}

\affiliation{Condensed Matter and Materials Science Department, Brookhaven National
Laboratory, New York, USA}

\author{F. von Rohr}

\affiliation{Department of Chemistry, Princeton University, New Jersey, USA}

\author{P. Pervan}

\affiliation{Institut za fiziku, Zagreb, Croatia}

\author{P. K. Das}

\affiliation{Istituto Officina dei Materiali (IOM-CNR), Laboratorio TASC, Trieste,
Italy}

\affiliation{International Centre for Theoretical Physics, Trieste, Italy}

\author{I. Vobornik}

\affiliation{Istituto Officina dei Materiali (IOM-CNR), Laboratorio TASC, Trieste,
Italy}

\author{R. J. Cava}

\affiliation{Department of Chemistry, Princeton University, New Jersey, USA}

\author{T. Valla}

\affiliation{Condensed Matter and Materials Science Department, Brookhaven National
Laboratory, New York, USA}

\maketitle
\onecolumngrid

\section*{supplemental material}

Figures S1 and S2 show the photon energy dependence of the ARPES maps
of the band structure of SnSe. Due to the loss of information about
the electron's perpendicular momentum ($k_{\perp}^{2}$) upon leaving
the surface, its value is estimated using the formula $\frac{\hbar^{2}}{2m_{e}}k_{\perp}^{2}=E_{kin}\cos^{2}\vartheta-V_{o}$
($\vartheta$ is the electron emission angle, 0 for $\bar{\Gamma})$
with only one free parameter—the inner potential $V_{o}$. The parameter
measures the position of the bottom of the free-electron final band
with respect to the vacuum level, and is usually found to be in the
range from -10 to -20~eV. Both values for the inner potential that
we used in Figures S1 and S2 confirm that the scans at six photon
energies (30-50 eV) sampled along three half-widths ($\frac{\pi}{a}=0.27\text{Å}^{-1}$)
of the bulk Brillouin zone. The dispersion along $k_{a}$ of the highest-lying
band has shown to be below our detection limit of 30–50~meV. 

Figure S3 shows in a different aspect ratio and color scale band dispersion
cuts of the pockets $w_{1}$ to $w_{4}$ along $\bar{\Gamma}-\bar{Z}$
and cuts of pockets $w_{4}$ and $v_{2}$ in the perpendicular direction
to justify the parabolic fitting of the bands on top of the valence
band. In addition to the fits of the maxima of photoemission intensity,
the latter two have been fitted by following half-width-at-half-maximum
points from above as well. 

The last section is describing our toy-model calculation of several
transport properties for a single band resembling the bands forming
the top of the valence band of SnSe. We argue that the slight mass
anisotropy cannot be accounted for the factor of 4 difference in conductivities
that has been observed along the $b$ and $c$ axes in SnSe.

\newpage{}
\begin{center}
\noindent\shadowbox{\begin{minipage}[t]{1\columnwidth - 2\fboxsep - 2\fboxrule - \shadowsize}%
\begin{center}
\includegraphics[width=160mm]{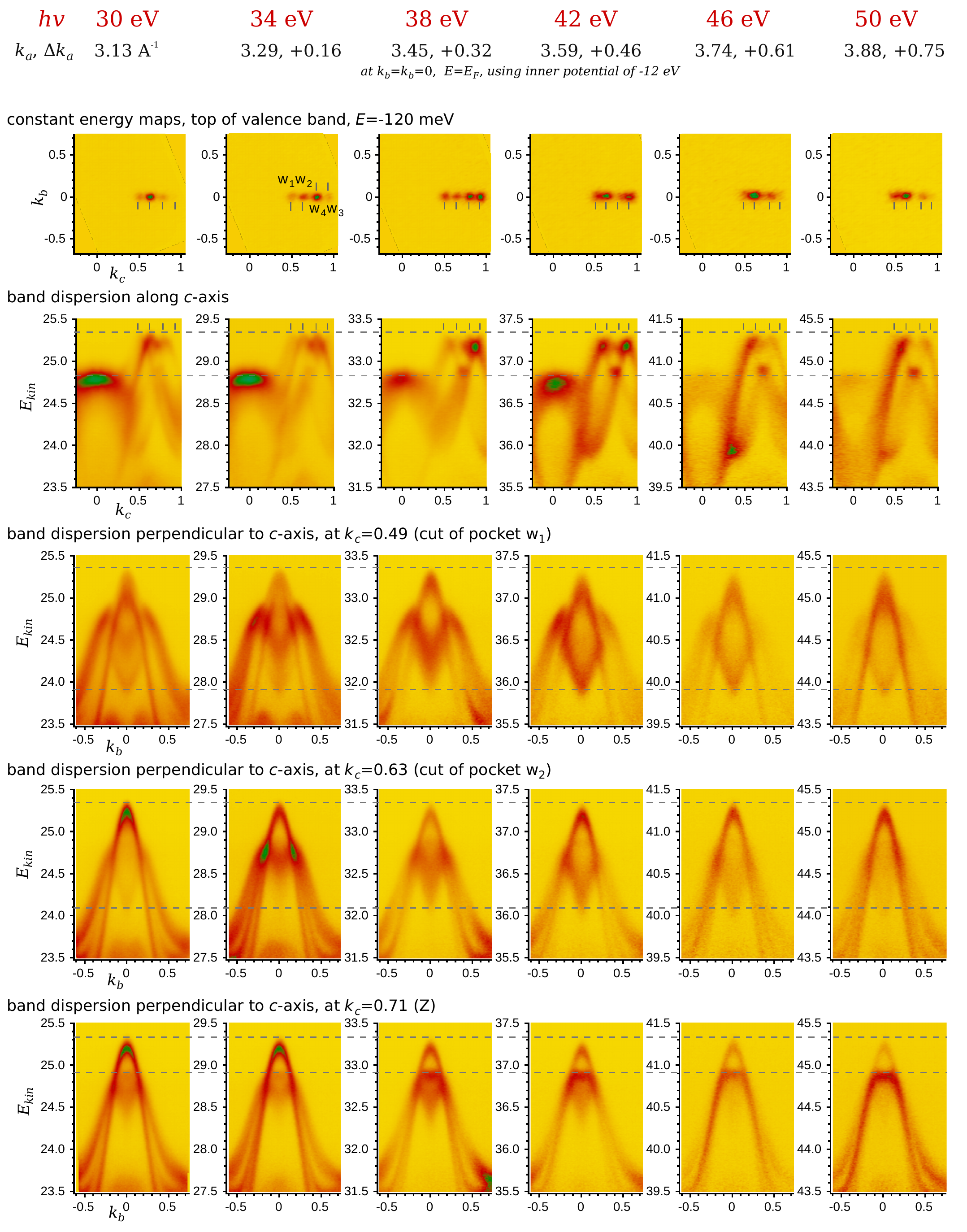}
\par\end{center}%
\end{minipage}}
\par\end{center}

\textbf{Figure S1}. Photon energy dependence of the ARPES maps of
the band structure of SnSe. Perpendicular momentum $k_{a}$ and its
relative change $\Delta k_{a}$ have been estimated using the inner
potential $V_{o}$ of -12~eV. Dashed horizontal lines are shown at
several characteristic energies as a guide to the eye.

\newpage{}
\begin{center}
\noindent\shadowbox{\begin{minipage}[t]{1\columnwidth - 2\fboxsep - 2\fboxrule - \shadowsize}%
\begin{center}
\includegraphics[width=160mm]{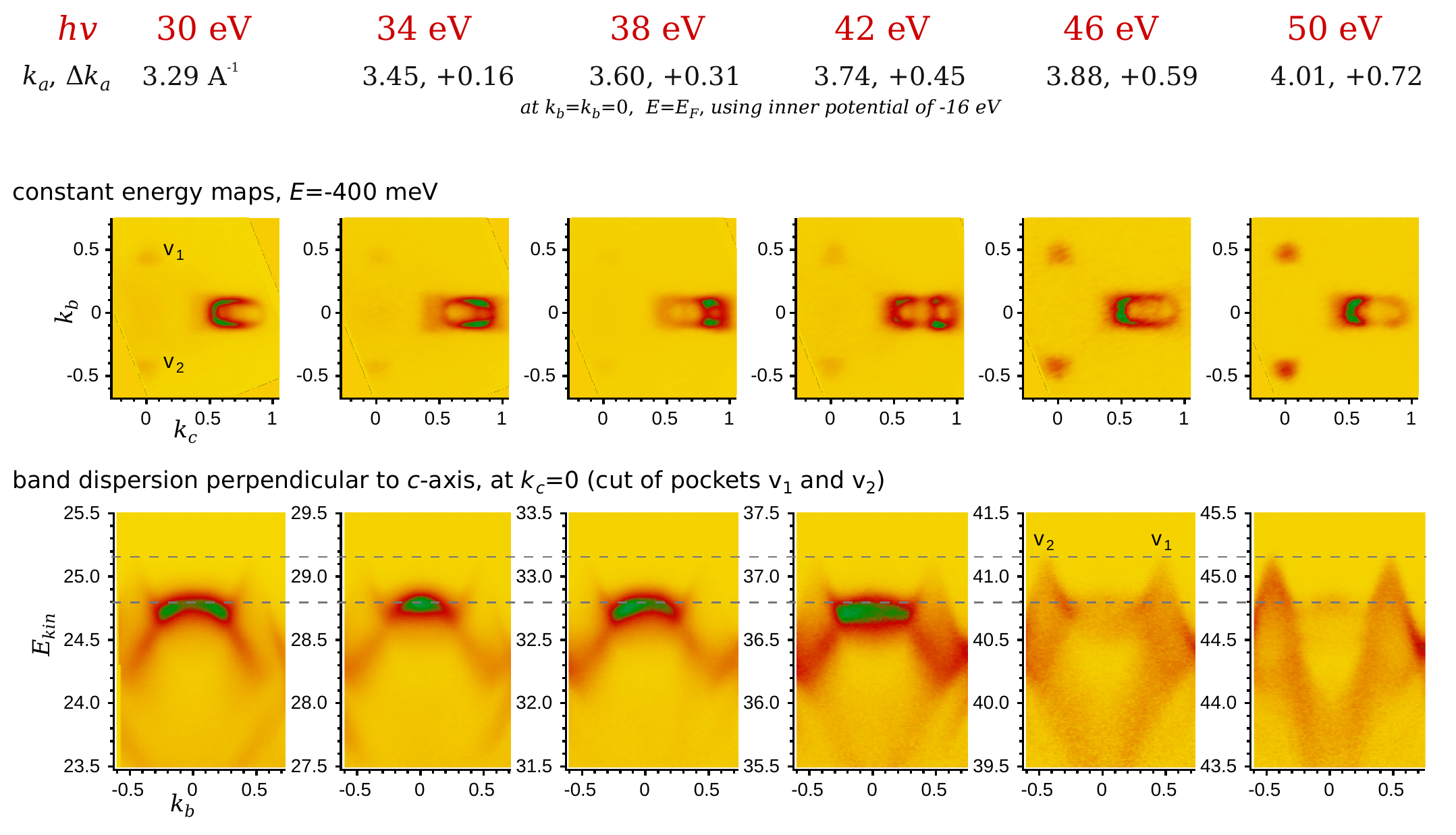}
\par\end{center}%
\end{minipage}}
\par\end{center}

\textbf{Figure S2}. Photon energy dependence of the ARPES maps of
the band structure of SnSe. Perpendicular momentum $k_{a}$ and its
relative change $\Delta k_{a}$ have been estimated using the inner
potential $V_{o}$ of -16~eV. Dashed horizontal lines are shown at
a few characteristic energies as a guide to the eye. 

\newpage{}
\begin{center}
\noindent\shadowbox{\begin{minipage}[t]{1\columnwidth - 2\fboxsep - 2\fboxrule - \shadowsize}%
\begin{center}
\includegraphics[width=160mm]{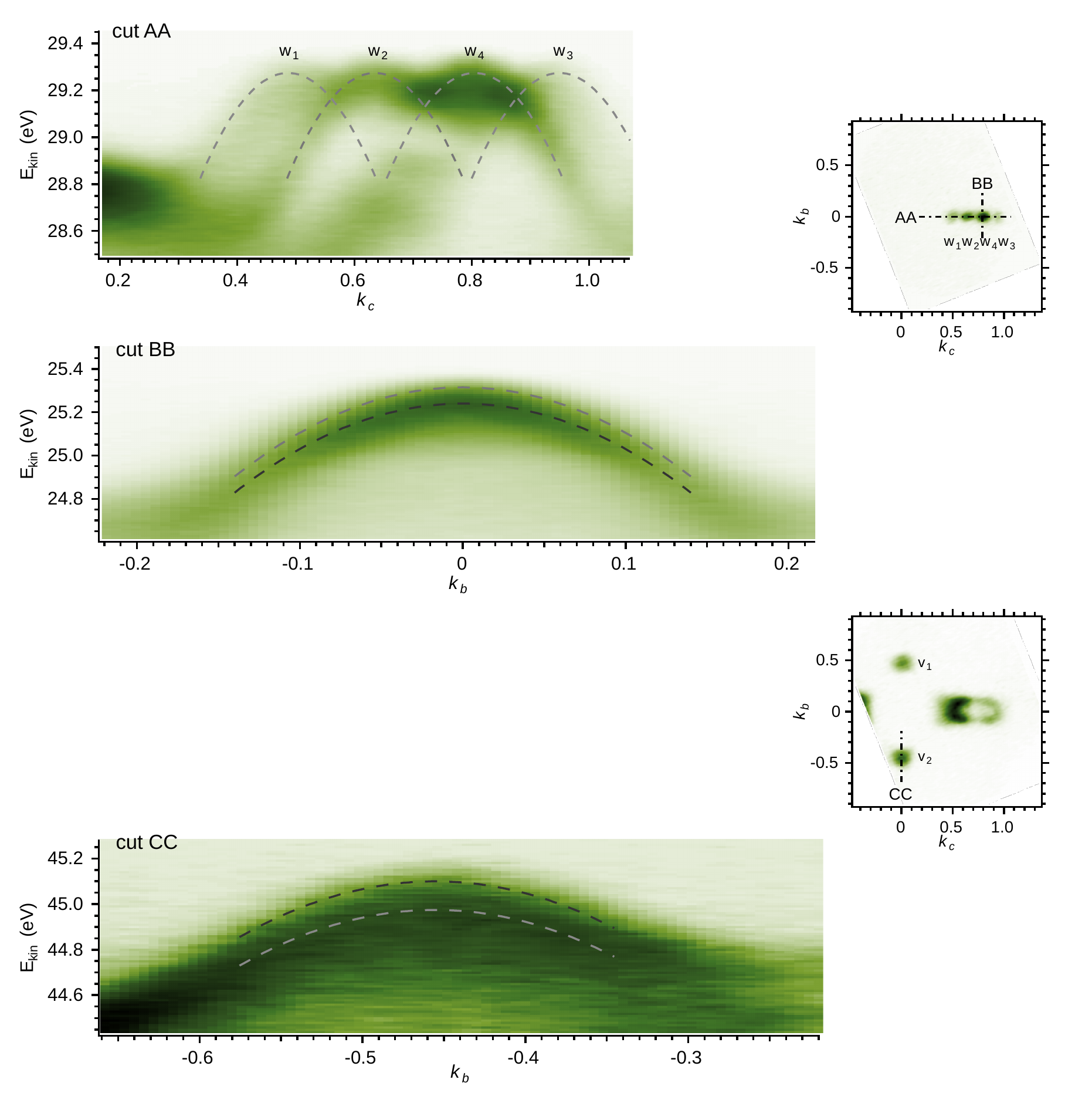}
\par\end{center}%
\end{minipage}}
\par\end{center}

\textbf{Figure S3}. Several band dispersion cuts of the pockets $w$
and $v$ shown in a different aspect ratio exemplifying the parabolic
shape of the bands. In addition to maximum-intensity parabolae (dashed
lines) overlaid to the cuts BB of the pocket $w_{4}$ and CC of the
pocket $v_{2}$ are parabolae following half-width-at-half-maximum
points from above. These are free from any intensity coming from neighboring
bands. Photons of 34, 30, and 50 eV were used, respectively. 

\newpage{}

\textbf{A toy-model calculation of conductivity and Seebeck coefficient
tensors} $\sigma$, $S$ for a single band with the in-plane parabolic
dispersion and a tight-binding-like dispersion across the layers,
mimicking the valence band of SnSe:
\[
\varepsilon(k_{a},k_{b},k_{c})=-\frac{\hbar^{2}k_{c}^{2}}{2m_{c}}-\frac{\hbar^{2}k_{b}^{2}}{2m_{b}}-\Delta(1-\cos ak_{a})
\]
in the Boltzmann transport equation formalism (González-Romero, arXiv:\href{http://arxiv.org/abs/1612.05967v1}{1612.05967})

\begin{minipage}[b][6cm][c]{10cm}%
\[
\sigma_{\alpha\beta}=-\frac{e^{2}}{(2\pi)^{3}}\iiint_{BZ}v_{\alpha}v_{\beta}\tau_{\vec{k}}\,\frac{\partial}{\partial\varepsilon}f_{o}(\varepsilon,\mu)\,d\vec{k}
\]

\[
(\sigma S)_{\alpha\beta}=-\frac{ek_{B}}{(2\pi)^{3}}\iiint_{BZ}v_{\alpha}v_{\beta}\tau_{\vec{k}}\,\frac{\varepsilon-\mu}{k_{B}T}\frac{\partial}{\partial\varepsilon}f_{o}(\varepsilon,\mu)\,d\vec{k}
\]

\begin{center}
$\alpha$ and $\beta$ denote the Cartesian axes 
\par\end{center}%
\end{minipage}%
\begin{minipage}[b][6cm][c]{6cm}%
\includegraphics[width=4.5cm]{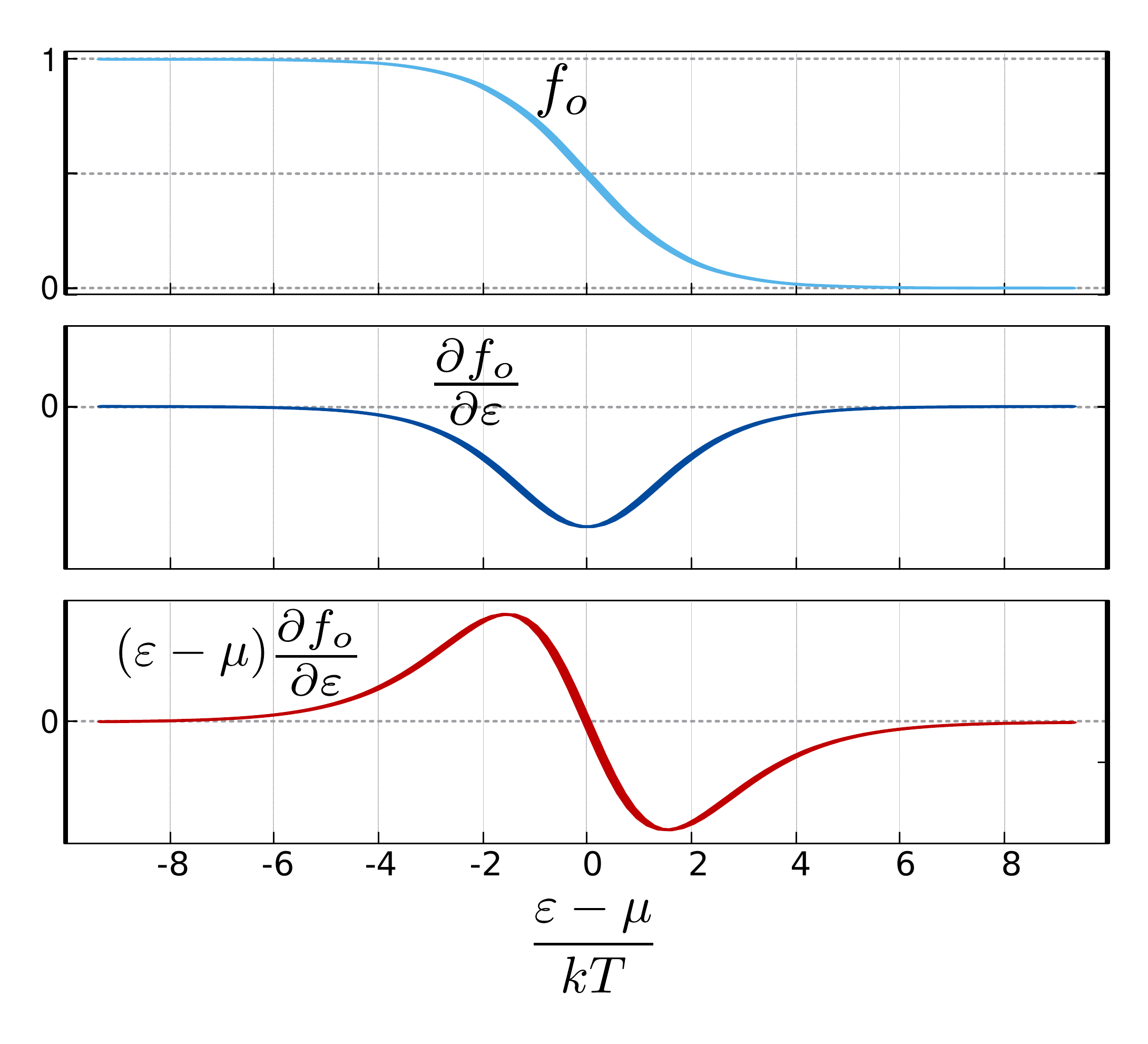}

$f_{o}(\varepsilon,\mu)=1/(\exp\frac{\varepsilon-\mu}{k_{B}T}+1)$

\medskip{}

$\frac{\partial}{\partial\varepsilon}f_{o}(\varepsilon,\mu)=\frac{1}{k_{B}T}f_{o}(f_{o}-1)$%
\end{minipage}

assuming an elliptical angular dependence of the in-plane scattering
time

\begin{minipage}[b][5cm][c]{7cm}%
\[
\tau(k_{b},k_{c})=\tau_{0}\sqrt{\frac{\tau_{b}^{2}k_{b}^{2}+\tau_{c}^{2}k_{c}^{2}}{k_{b}^{2}+k_{c}^{2}}}
\]
\end{minipage}%
\begin{minipage}[b][5cm][c]{8cm}%
\includegraphics[width=4.5cm]{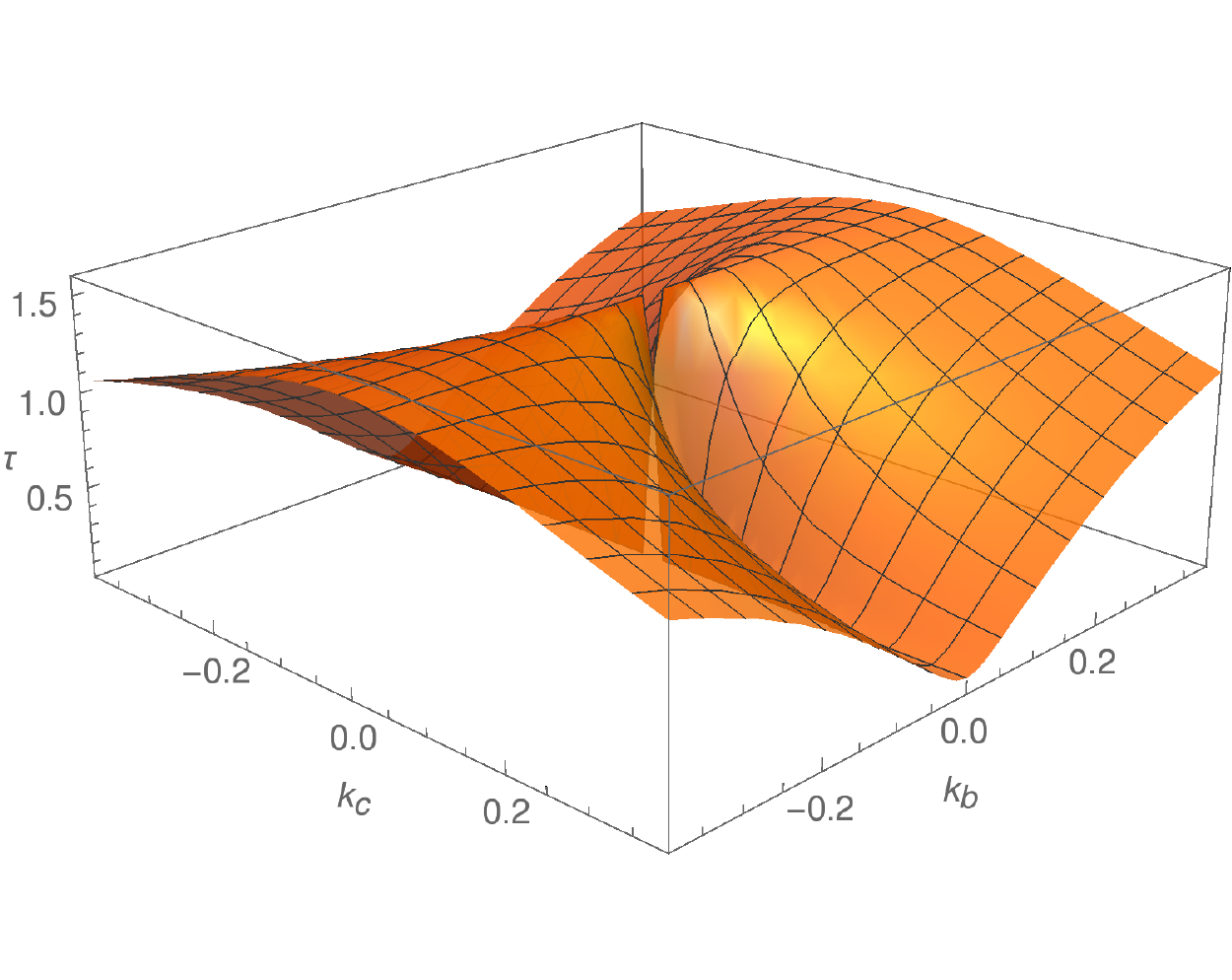}\includegraphics[width=3.5cm]{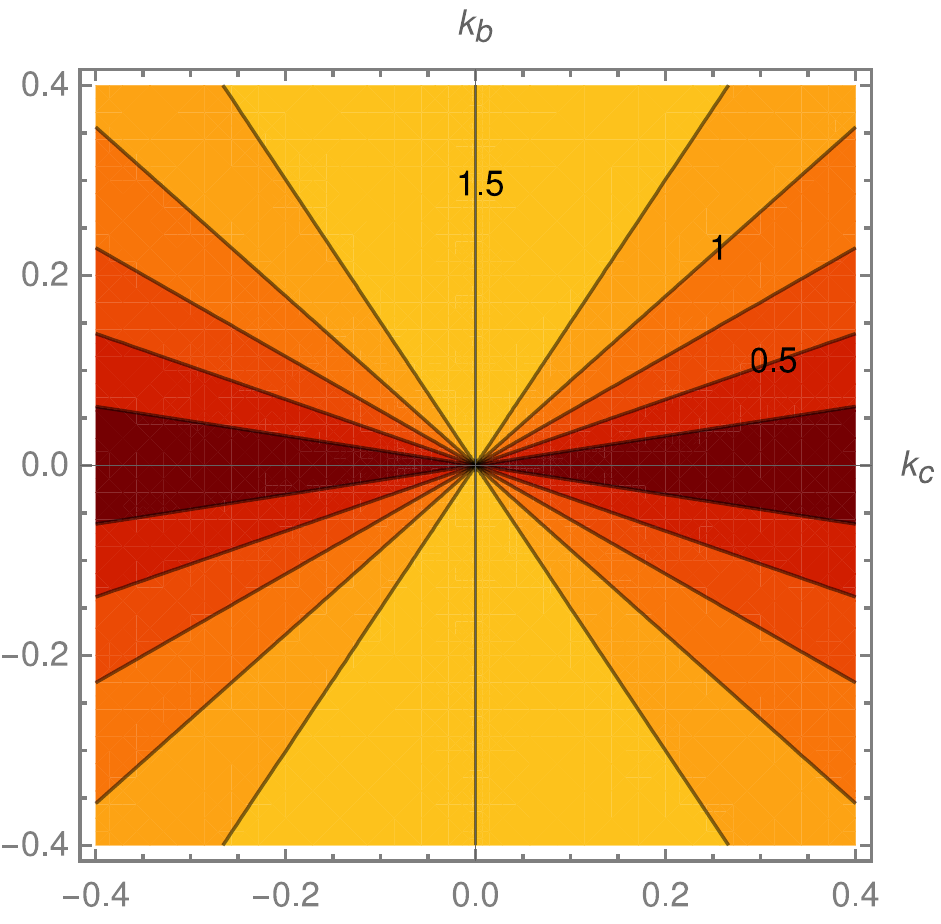}%
\end{minipage}

Here, $v_{a}=-\frac{\Delta}{\hbar}a\sin ak_{a}$, $v_{b}=-\frac{\hbar}{m_{b}}k_{b}$,
$v_{c}=-\frac{\hbar}{m_{c}}k_{c}$. Scattering time $\tau_{0}$ is
usually taken to be of the order of several femtoseconds. The effective
masses are nearly isotropic, $m_{c}=0.21$, $m_{b}=0.18$. $k_{B}T=25$
meV.

\begin{minipage}[t][7\totalheight]{7cm}%
Setting the ratio $\tau_{b}/\tau_{c}$ to 15, this simple model gives
$\sigma_{bb}/\sigma_{cc}=2.5$. The anisotropy found in angle-resolved
transport measurements is as high as 4 (Xu \emph{et al}, DOI:\href{https://doi.org/10.1021/acsami.7b00782}{10.1021/acsami.7b00782}).

Three values of $2\Delta$, the total band dispersion across the layers,
were used: $50$ meV, $100$ meV (lighter curves), and $25$ meV (darker
curves).

Interestingly, the in-plane Seebeck coefficient tensor components
are insensitive to the scattering time anisotropy, and the power factor
is only affected in the first order, through $\sigma$.

Smaller band width in the perpendicular direction, i.e. weak bonding
between the layers, leads to higher conductivities and power factors
for a given chemical potential, but has negligible influence on the
Seebeck coefficient.%
\end{minipage}\hspace{0.5cm}%
\begin{minipage}[t][7\totalheight]{6cm}%
\includegraphics[width=6cm]{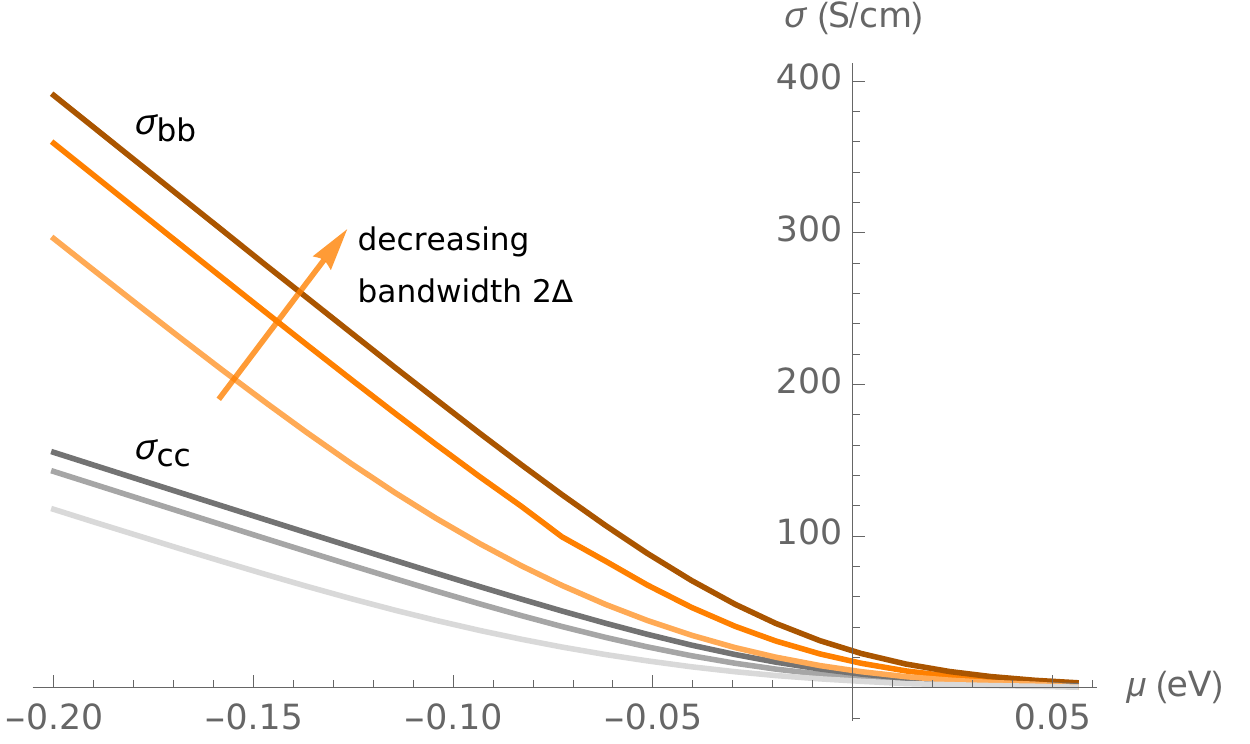}

\includegraphics[width=6cm]{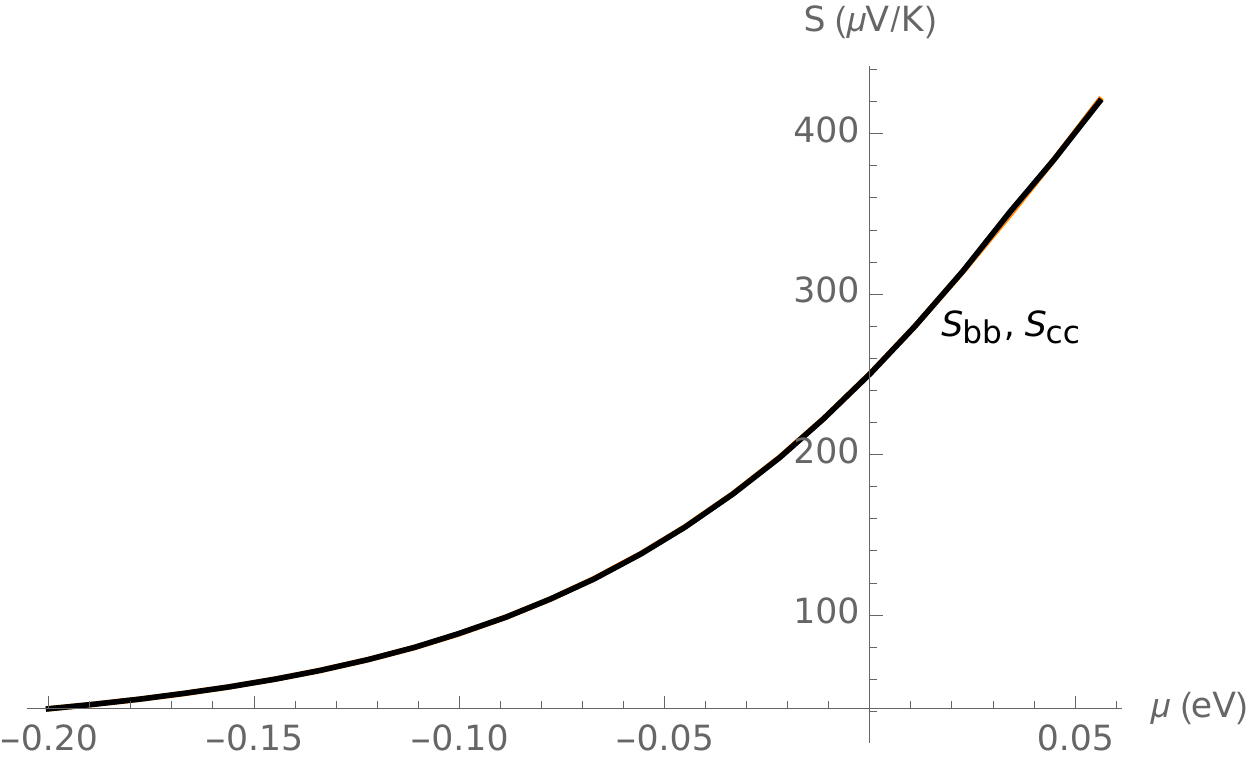}

\includegraphics[width=6cm]{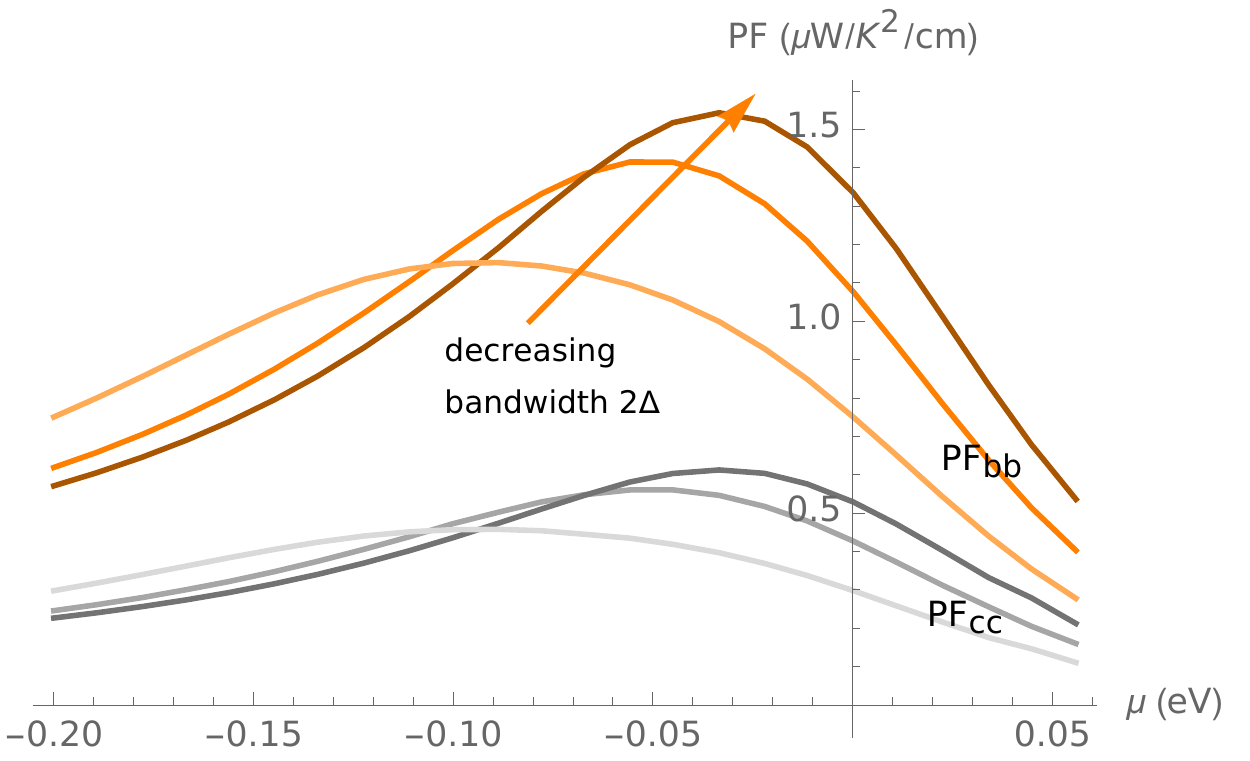}%
\end{minipage}

\end{document}